\documentclass[pra, twocolumn, showpacs]{revtex4}

\usepackage{amsfonts}
\usepackage{amssymb}
\usepackage[dvips]{graphicx}
\usepackage{amsmath}

\newcommand{\ket}[1]    {| #1 \rangle}

\newcommand{\Tin}{T}
\newcommand{\Tout}{50\%}
\renewcommand{\t}[1]{\textrm{#1}}
\newcommand{\mean}[1]{\bar #1}

\begin{document}
\title{Multi-pass classical vs. quantum strategies in lossy phase estimation}

\date{\today}
\pacs{03.65.Ta, 06.20.Dk, 42.50.St}
\author{Rafa{\l} Demkowicz-Dobrza{\'n}ski\footnote{demko@fizyka.umk.pl}}
\affiliation{Institute of Physics, Nicolaus Copernicus University, ul.~Grudziadzka 5, PL-87-100 Toru\'{n}, Poland}
\begin{abstract}
The use of classical multiple-pass approach for phase estimation which mimics the behavior of the N00N states,
is compared with quantum techniques. It is shown that in the presence of
losses its performance is significantly worse than the one of the optimal quantum strategy.
\end{abstract}
\maketitle
\section{Introduction}
Optical interferometry \cite{Hariharan2003} is capable of ultra precise measurement of various physical quantities
such as length, time, temperature, velocity provided they can be translated into a phase shift $\varphi$
of the light beam. In an interferometric experiment, light intensities detected at the output ports
carry information on the \emph{true} value of $\varphi$. Once intensities are measured the estimation procedure
may be carried out resulting in an \emph{estimated} value $\tilde{\varphi}$. The intrinsic character of the detection process, however,
is stochastic. Within the semi-classical approach this is attributed to the quantum nature of atoms constituting the detectors,
which under the interaction with classical light field are being excited. The Poissonian statistics governing
the number of excited atoms, and therefore the number of detector clicks -- \emph{the shot noise} -- leads to
phase estimation uncertainty scaling: $\delta \varphi \propto 1/\sqrt{\mean{n}}$ where $\mean{n}$ is the mean number of detector clicks.

Quantum description of light, provided by quantum optics, allows for a deeper analysis of what are the ultimate
limits of phase estimation precision. Semi-classical $\delta \varphi \propto 1/\sqrt{\mean{n}}$ scaling
can be recovered when one restricts oneself to the use of coherent states of light,
and $\mean{n}$ is identified with the mean number of photons used. What is more interesting, however,
is that other inherently quantum states of light, such as e.g.
squeezed states, twin-Fock, N00N states, offer in certain scenarios quadratic improvement of phase estimation precision
leading to \emph{the Heisenberg limit} (HL): $\delta \varphi \propto 1/\mean{n}$
\cite{Giovannetti2004, Caves1981, Grangier1987, Xiao1987, Holland1993, Sanders2003, Bolinger1996, Dowling1998}.
The quantum-enhanced precision comes at a price. Quantum states achieving the HL require highly nontrivial preparation procedures,
such as strong squeezing or large entanglement. Moreover, the ultra-high phase sensitivity of the states is
usually accompanied by low robustness to environmental decoherence. In particular, photon loss renders the N00N states useless and
other more robust classes of quantum states needs to be employed \cite{Huelga1997, Shaji2007, Sarovar2006, Rubin2007, Gilbert2008, Huver2008, Dorner2008, Demkowicz2009a, Kacprowicz2009}.
Fragility of quantum states and difficulties in their preparation
are the challenges that need to be met if quantum enhanced interferometry is to be implemented in practice.

The situation prompted ideas on overcoming the shot noise limit without resorting to the use of sophisticated quantum states,
but rather allowing some minor modifications in interferometric experiment. One of the most simple
techniques is the multi-pass approach, in which a light beam is allowed to pass the sample many times and thus acquire
a multiple of the phase to be estimated -- in this way it mimics the behavior of the N00N state
 \cite{Higgins2007, Higgins2009}. Similar ideas were also analyzed in the context of precise clock synchronization \cite{Burgh2005, Boixo2006}.
Obviously such an approach can be seen as a change in the rules of the game, and the use
of quantum states may be still argued to be the only way to reach the HL.

The presence of losses, complicates the relation between classically based and quantum based strategies even further,
neither N00N nor multi-pass approach is the optimal strategy. In what follows the optimal use of classical states for
interferometry in the presence of losses is investigated and compared with the optimal use of quantum states.
It is shown that no straightforward
classical multi-pass technique can reach the precision offered by quantum states when losses are present.

\begin{figure}[h]
\includegraphics[width= \columnwidth]{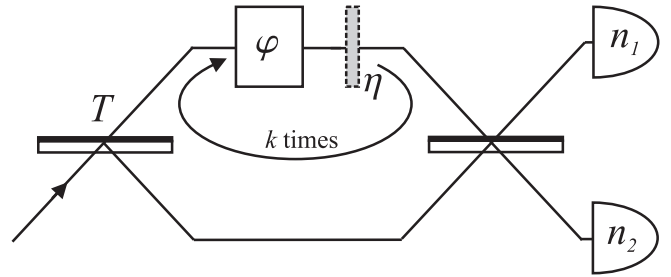}
\caption{The generic Mach-Zennder interferometer fed with a classical light beam. The phase shift $\varphi$ inside the interferometer modulates the output intensities. Possible losses accompanying the phase shift are represented by a beam splitter with power transmission $\eta$.
In multi-pass strategies light passes the sample $k$ times.}
\label{fig:mz}
\end{figure}
\section{Classical interferometry in the presence of loss}
A paradigmatic model for the study of phase estimation is the Mach-Zehnder interferometer (MZI) depicted in Fig.~\ref{fig:mz}.
A classical light beam is fed into the lower input port of the interferometer.
The beam is divided at the input beam-splitter (transmissivity $\Tin$),
experiences a relative phase delay $\varphi$ caused by a phase shift element placed in the upper arm,
and---if no multi-passes are allowed---is recombined at the output beam-splitter (transmissivity $\Tout$).
Finally, its intensity is being measured at the two output ports.
The phase shift element, apart from introducing a phase delay, typically is also a source of losses.
This can be modeled by placing an additional beam splitter with transmissivity $\eta$ in the upper arm.
A change of $\varphi$ can be sensed by observing the change of light intensities at the output ports.

The probability that the two detectors register $n_1$ and $n_2$ clicks, provided the unknown phase shift equals $\varphi$ can be expressed
as a product of two Poisson distributions $p(n_1,n_2|\varphi) = P_{\mean{n_1}}(n_1) P_{\mean{n_2}}(n_2)$, with respective mean values
\begin{equation}
\mean{n_1} = A\mean{n} \left(1 - v \cos \varphi\right), \quad \mean{n_2} = A \mean{n} \left(1 + v \cos \varphi \right)
\end{equation}
where $\mean{n}$ is the mean number of photons in the input beam, while:
\begin{equation}
A=\frac{1-\Tin(1-\eta)}{2}, \quad v=\frac{2 \sqrt{\Tin(1-\Tin)\eta}}{1-\Tin(1-\eta)}.
\end{equation}
For $\eta=1$ (lossless case), the choice $\Tin=50\%$ causes the functions $\mean{n_1}(\varphi)$, $\mean{n_2}(\varphi)$ to enjoy both the
$v=100\%$ visibility and the maximum amplitude $A=\mean{n}/2$. Any other value of $\Tin$ will lead to a decreased visibility and hence
it is no surprise that $\Tin = 50\%$ is the optimal choice for an interferometric experiment.
For $\eta<1$, however, the optimal choice for $\Tin$ is no longer obvious and one needs more
precise tools to find the solution.

\begin{figure}[t]
\includegraphics[width=0.45 \textwidth]{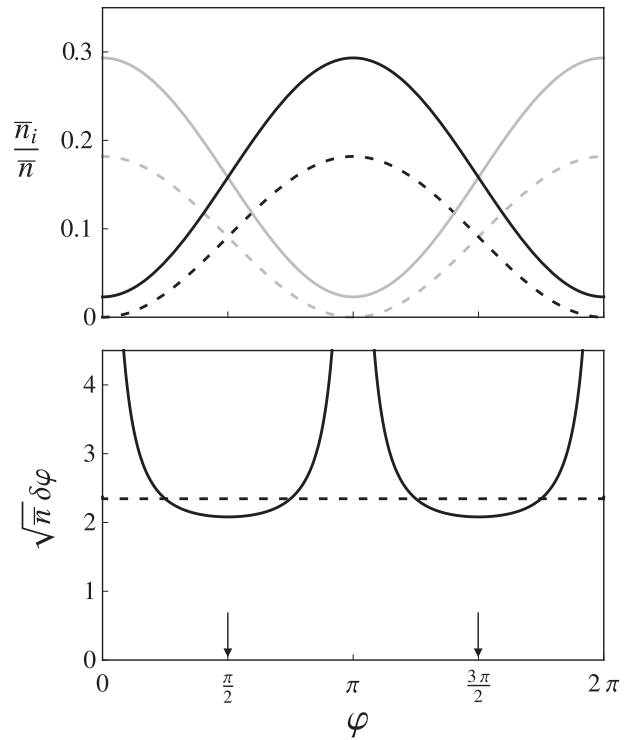}
\caption{Mean number of clicks registered by detectors (top) for $\eta=0.1$ when the transmission of the input beam
splitter of Mach-Zehnder interferometer
is chosen to maximize visibility (dashed), or to minimize estimation uncertainty (solid).
The dependence of uncertainty on $\varphi$ (bottom) illustrates the fact that the maximum-visibility strategy
works equally well for the whole $2\pi$ region, whereas the optimal strategy requires operating in the vicinity of phases indicated by arrows, where it achieves the standard interferometric limit.}
\label{fig:fishvis}
\end{figure}

Knowledge of $p(n_1,n_2|\varphi)$, and measurement results $n_1$, $n_2$ allows for
an estimation of the value of $\varphi$. One of the key tools in estimation theory is the Cram{\'e}r-Rao inequality \cite{Helstrom1976}
which provides a fundamental bound for the root mean square error $\delta \varphi$
of estimation expressed in terms of the Fisher information $F$:
\begin{equation}
\label{eq:fisher}
\delta\varphi \geq \delta\varphi_{\t{min}} = \frac{1}{\sqrt{F}}, \ F=\sum_{n_1,n_2} \frac{1}{p(n_1,n_2|\varphi)} \left(\frac{\partial p(n_1,n_2|\varphi)}{\partial \varphi}\right)^2.
\end{equation}
Loosely speaking, the larger is the derivative of $p(n_1,n_2|\varphi)$ over $\varphi$, the larger is the
Fisher information, the lower is the achievable variance of estimation and consequently the
better is the estimation precision.
For the setup considered the Fisher information reads explicitly:
\begin{equation}
F=\frac{4 \mean{n} A v^2 \sin^2\varphi }{2 - v^2 (\cos 2 \varphi + 1)}.
\end{equation}
It is intuitive to expect that in the presence of losses ($\eta <1$), the optimal choice for $\Tin$ should
correspond to more light being transmitted into the lossy arm in order to keep the visibility of interference high.
Choosing $\Tin=1/(1+\eta)$ we retain $100\%$ visibility ($v=1$), and get $F=2 \mean{n} \eta/(1+\eta)$.
Surprisingly, this is not the optimal choice. The optimal choice $\Tin=1/(1+\sqrt{\eta})$ results from
a compromise between keeping visibility high and loosing as little light as possible, and yields:
$F=4 \mean{n} \eta \sin^2\varphi /(1+ \eta -2 \sqrt{\eta} \cos 2\varphi)$. The price to pay is the dependence of $F$ on the value of $\varphi$,
and the optimality is achieved only in the most sensitive estimation region corresponding to $\varphi=\pi/2, 3\pi/2$, where
 $\mean{n_1}(\varphi)$, $\mean{n_2}(\varphi)$ curves are the steepest.
In this region, we get $F=4 \mean{n} \eta/(1+\sqrt{\eta})^2 $
and the minimal uncertainty:
\begin{equation}
\delta\varphi_{\t{min}}^{\t{SIL}}  = \frac{1+\sqrt{\eta}}{2 \sqrt{\mean{n} \eta}}.
\end{equation}
The above classical precision obtained for the optimal choice of beam splitter transmission is referred as the
\emph{standard interferometric limit} (SIL) \cite{Dorner2008, Demkowicz2009a}, with its characteristic $1/\sqrt{\mean{n}}$ shot-noise scaling.
Fig.~\ref{fig:fishvis} depicts the differences between the optimal and the maximum-visibility estimation strategy.

The intuitive argument: the steeper the curves
 the better the precision, should be taken with care. In particular for the $100\%$ visibility case, Fisher information is
 constant over the whole $[0,2\pi]$ region even though the steepness of the curves varies greatly.
 When the curves become flat, one would expect vanishing of the Fisher information. This is not the case, however, since
  simultaneously one of the mean values $\mean{n_1}(\varphi)$, $\mean{n_2}(\varphi)$ approaches zero, and this causes
 the $1/p(n_1,n_2|\varphi)$ term to blow up keeping the Fisher information constant.

\section{Quantum strategies}
Optimal quantum strategies allow for a significant improvement over the SIL even in the presence of losses
\cite{Huver2008, Dorner2008, Demkowicz2009a}.
For the sake of generality one replaces the input beam-splitter of MZI with a general quantum
state preparation device, and analogously the output beam-splitter plus detectors with a general quantum measurement.
In the absence of losses ($\eta=1$), the optimal $n$ photon state that should be prepared by the preparation device
is the N00N state \cite{Bolinger1996} $\ket{\psi}=(\ket{n}\otimes \ket{0} + \ket{0}\otimes {\ket{n}})/\sqrt{2}$, where the first (second)
factor in the tensor product corresponds to the number of photons in the upper (lower) arm of the interferometer.
The minimal estimation uncertainty corresponds to the HL:
\begin{equation}
\delta \varphi_{\t{min}}^{\t{HL}} = \frac{1}{n}
\end{equation}
which is a quadratic improvement over SIL.
The improvement stems from the fact that the two superposition terms acquire a relative phase factor
$\exp(i n \varphi)$, which causes measured fringes to be $n$ times denser (and hence $n$ times steeper) than
if a single photon or classical light were used.
In the presence of losses, however, NOON state performance deteriorates rapidly. Even, if we consider a N00N state
with optimally distributed weights among both superposition terms the increase of uncertainty is exponential in $n$:
\begin{equation}
\delta \varphi_{\t{min}}^{\t{N00N}} = \frac{1 + \eta^{n/2}}{2 n \eta^{n/2}}
\end{equation}
and for large $n$ its advantage over classical strategies is quickly washed out even in the presence of moderate losses.
Exponential growth of uncertainty may be avoided if
instead of a single $n$ photon N00N state, a $k$ photon N00N state is sent $m=n/k$ times through the interferometer
\cite{Dorner2008, Demkowicz2009a}
\begin{equation}
\label{eq:noonchop}
\delta \varphi_{\t{min}}^{\t{CHOP}} = \frac{1 + \eta^{k/2}}{2 k \eta^{k/2} \sqrt{m}}.
\end{equation}
Relaxing the integer constraint on $m$ and $k$, the optimization over the choice of $k \in [1,n]$ yields
\begin{equation}
\delta\varphi_{\t{min}}^{\t{CHOP}}=
\begin{cases}
  \frac{1+\sqrt{\eta}}{2\sqrt{n\eta}}& ;\eta\le \eta_0 \approx 0.228 \\
  \frac{1+\sqrt{\eta_0}}{2\sqrt{n\eta_0}}\sqrt{\frac{\ln{\eta}}{\ln{\eta_0}}}&  ;\eta_0 <\eta\le \eta_0^{\frac{1}{n}}\\
  \frac{1+\eta^{\frac{n}{2}}}{2 n\eta^{\frac{n}{2}}}& ;\eta >
  \eta_0^{\frac{1}{n}}
\end{cases}
\end{equation}
where the optimal choice of $k$ for the three regimes indicated above
is $k=1$, $k=1.478/|\ln{\eta}|$ (i.e. the solution of $1+\eta^{k/2}+
k\ln\eta=0$) and $k=n$.

More robust states have been proposed \cite{Dorner2008, Demkowicz2009a}
which outperform both classical and N00N state based strategies and raise hopes for
 beating SIL in realistic environments. The states are found by numerical optimization of
 quantum Fisher information which unlike its classical counterpart in Eq.~(\ref{eq:fisher}), does not require
 specifying the actual measurement performed on quantum states.
 For a general state of $n$ photons distributed among the two arm of the interferometer
 \begin{equation}
\ket{\psi} = \sum_{s=0}^n \alpha_s  \ket{s} \otimes \ket{n-s}
\end{equation}
phase estimation uncertainty reads:
 \begin{equation}
 \label{eq:qfisher}
 \delta \varphi_{\t{min}}^Q= \frac{1}{2} \left(\sum_{s=0}^n s^2 x_s - \sum_{l=0}^n\frac{\left[ \sum_{s=l}^{n} x_s s B^s_l(\eta) \right]^2}{\sum_{s=l}^{n} x_s B^s_{l}(\eta)}\right)^{-1/2},
 \end{equation}
where $x_s=|\alpha_s|^2$, $B^s_l(\eta) = \binom{k}{l}\eta^{s-l}(1-\eta)^{l}$. Optimal states result from
minimization of $\delta \varphi_{\t{min}}^Q$ over $x_s$.
 These states have complex structure and their optimality is the result of a subtle interplay between
the need for quantum enhanced phase sensitivity and robustness to losses.
While the estimation uncertainty is greatly reduced when using these states for realistic system parameters
the HL cannot be retained \cite{Dorner2008, Demkowicz2009a}.

\section{Multiple-pass classical strategies}
Comparing different estimation strategies requires a precise definition of resources that are available.
In most application the mean number of photons used is a relevant quantity.
Other approaches are conceivable however. One could fix the mean photon number \emph{traveling
through the sample}, which is relevant in situations when not the total power consumed is the limitation but
only the fraction of it that is absorbed by the sample.
One of the examples in which this distinction is relevant is the multi-pass approach to phase estimation.
Instead of sending light only once through the sample, one may reflect it and make it pass the sample
$k$ times, acquiring a multiple $k \varphi$ of the estimated phase.
This simple trick mimics the behavior of N00N states, making interference fringes $k$ times denser
and therefore allowing for better phase sensitivity.
According to Eq.~(\ref{eq:fisher}), Fisher information depends quadratically on the derivative of
the probability distribution over phase. Increasing the density of fringes $k$ times
makes the Fisher information grow as $k^2$, and consequently results
in $1/k$ scaling $\delta \varphi_{min}$.

Two separate cases may be considered: when multiple-passes are for free or when they are included in the overall budget as
as a resource equivalent to photons.

\subsection{Multiple-passes as a resource}
If the total resources are to be fixed to $\mean{n}$, the intensity of light passing through the sample $k$ times
has to be reduced to $\mean{n}/k$. This way we obtain interference fringes which are $k$ times denser,
at the cost of replacing $\mean{n}$ with $\mean{n}/k$ and $\eta$ with $\eta^k$. Mathematically, this strategy
is equivalent to the N00N chopping strategy with $k$ photon N00N states sent through the interferometer $\mean{n}/k$ times,
resulting in Eqs.~(\ref{eq:noonchop}). In the context of clock synchronization a analogous
equivalence in susceptibility to external decoherence has been observed between quantum GHZ states and multi-pass strategies using single qubit states
\cite{Burgh2005, Boixo2006}.

\subsection{Multiple-passes for free}
If the number of bounces $k$ does not contribute to the consumed resources, we can choose the value of $k$
that yields the highest sensitivity of estimation.
Mathematically, this situation is equivalent to replacing $\varphi \rightarrow k \varphi$, $\eta \rightarrow \eta^k$.
Choosing the optimal $\Tin=1/(1 +\eta^{k/2})$ in he MZI we get the estimation uncertainty  bounded by:
\begin{equation}
\delta \varphi_{\t{min}} = \frac{1+\eta^{k/2}}{2 k \sqrt{\mean{n} \eta^k}}.
\end{equation}
Notice that in the absence of losses the precision can be arbitrary good since the bigger the $k$ the lower the uncertainty.
When losses are present, minimal $\delta \varphi_{min}$ is achieved for $k= -2(1+\xi)/\ln \eta$,
where $\xi \approx 0.278$ is the solution to $\xi e^{\xi +1}=1$. The resulting uncertainty:
\begin{equation}
\delta \varphi_{\t{min}}^{\t{MP}} = \frac{\ln \eta}{4 \sqrt{\mean{n}} \xi}.
\end{equation}

\section{Multiple-pass quantum strategies}
For the full picture we also need to investigate the usefulness of multi-pass approach when employing quantum states.

\subsection{Multiple-passes as a resource}
Neglecting losses for the moment, consider a general $n$ photon quantum state which has experienced a phase shift $\varphi$ in the upper arm with respect to the lower arm of the interferometer:
\begin{equation}
\label{eq:qstate}
\ket{\psi} = \sum_{s=0}^n \alpha_s e^{- i s \varphi} \ket{s} \otimes \ket{n-s}.
\end{equation}
Multiple-passes give the advantage of multiple acquisition of phase and hence better sensitivity.
If the light passes the sample $k$ times, we need to replace $e^{- i s \varphi}$ with $e^{- i k s \varphi}$.
The same effect, however, can be obtained with a single pass of $k s$ photons
prepared in state:
\begin{equation}
\ket{\psi} = \sum_{s=0}^n \alpha_s e^{- i k s \varphi} \ket{k s} \otimes \ket{k n -k s}.
\end{equation}
Therefore, there is \emph{no advantage}, except for technical difficulties in preparing multi-photon quantum states, in using multiple-pass approach if multiple-passes are treated as a resource equivalent to photons. The same conclusion holds also in the presence of losses, since
the probability of loss behaves in exactly the same way, irrespectively whether a given number of photons pass $k$  times through the sample or
a $k$-multiple of them pass through the sample once.

\subsection{Multiple-passes for free}
The above argument clearly cannot be applied in the case when we may freely choose the number of passes and do not account for them in
the total amount of resources consumed. This approach will in general improve estimation precision as compared with
the optimal quantum single-pass strategy. When the general $n$ photon state given in Eq.~(\ref{eq:qstate})
passes the sample $k$ times, we can similarly as in the single pass case derive the formula for minimal uncertainty
and the only difference in the final result is the replacement of $\eta$ by $\eta^k$ and the division of the
whole formula by $k$:
 \begin{equation}
 \label{eq:qfisher}
 \delta \varphi_{\t{min}}^{\t{QMP}}= \frac{1}{2k} \left(\sum_{s=0}^n s^2 x_s - \sum_{l=0}^n\frac{\left[ \sum_{s=l}^{n} x_s s B^s_l(\eta^k) \right]^2}{\sum_{s=l}^{n} x_s B^s_{l}(\eta^k)}\right)^{-1/2}.
 \end{equation}
The optimal strategy can be found by direct minimization of $\delta \varphi_{min}^{\t{QMP}}$
over $x_s$ and $k$ parameters. Due to the additional parameter
$k$ which enters the formula in a nontrivial way, numerical optimization is a bit more complicated than in a single pass case.
Unlike $ \delta \varphi_{min}^{Q}$ from Eq.~(\ref{eq:qfisher}) which was proven a convex function of $x_s$ \cite{Demkowicz2009a},
$\delta \varphi_{min}^{\t{QMP}}$
is not a convex function of $k$. One needs to take an additional care not to end up in a local minimum instead of a global one.

\section{Comparison}
\begin{figure}
\includegraphics[width= \columnwidth]{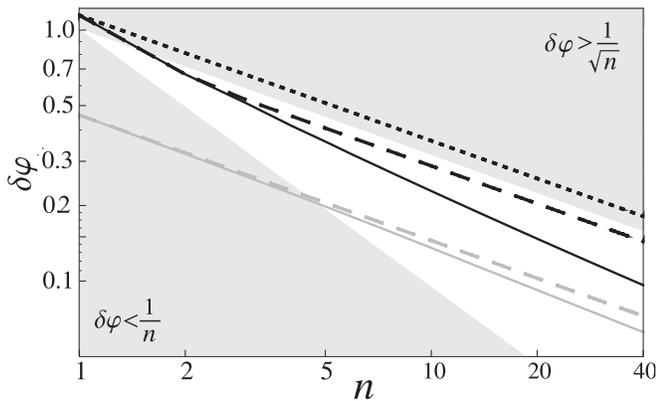}
\caption{Log-log scale plot of phase estimation uncertainty for transmission $\eta=0.6$
 for optimal classical single-pass strategy (black, dotted), quantum N00N chopping strategy or equivalently
classical multi-pass strategy (black, dashed), optimal quantum strategy (black, solid),
where $n$ denotes total resources consumed equal to the number of photons used times number of passes through the sample.
If multiple passes are not considered a resource, the uncertainty decreases, which
is depicted by two gray curves, representing optimal classical multi-pass strategy (gray, dashed) and quantum multi-pass strategy (gray, solid).
Lower shaded area corresponds to uncertainties smaller than that given by the Heisenberg limit: $1/n$, while the upper shaded area
corresponds to uncertainties larger than the shot noise limit $1/\sqrt{n}$.
}
\label{fig:logplots}
\end{figure}

Comparison of estimation uncertainties achievable using all strategies presented in the paper for a particular value of interferometer
transmission $\eta=0.6$ is depicted in Fig.~\ref{fig:logplots} as a function of number of photons used.
Quantum single-pass strategy clearly outperforms classical single-pass strategy as well as classical multi-pass strategy where
multiple-passes are treated as a resource. Notice the $1/\sqrt{n}$ character of the curves for classical strategies, as opposed
to a steeper one for the quantum. More importantly, with growing number $n$ the advantage of the optimal quantum strategy
is more evident. Nevertheless, due to losses it is not powerful enough to
 provide $1/n$ HL  which is achievable with N00N states in the absence of losses.

When multiple-passes are allowed for free, the uncertainties of both the optimal classical and quantum strategies
are significantly decreased, falling for small $n$ even below the HL. Notice however, that
the slope of the classical curve manifests again its $1/\sqrt{n}$ character. Quantum curve is again steeper and
proves its advantage over classical one more and more the bigger $n$ becomes.

\section{Summary}
This paper analyzed in detail the impact of  multiple-passes for phase estimation
using both classical and quantum strategies. In particular it has been shown that while
in the absence of losses classical multi-pass strategy may in some sense be regarded as equivalent to
the optimal quantum strategies this is no longer the case when losses are present.
It might sound as a paradox, since we are used to the fact that quantum states
manifest their advantage over classical ones best in well controlled environment with low noise.
In the phase estimation problems, however, if multiple-pass approach is allowed then,
indeed it is \emph{only} in the presence of losses that quantum states retain their advantage over classical
based strategies in the quest for precise phase estimation.

\acknowledgments
I am grateful to Konrad Banaszek for many fruitful discussions and constant support.
This research was supported by the Future and Emerging Technologies (FET) programme within the Seventh
Framework Programme for Research of the European Commission, under the FET-Open grant agreement CORNER no.\  FP7-ICT-213681
and the Polish Ministry of Science and Higher Education under grant no.\ N~N202~1489~33.

\end{document}